\documentclass[sigconf,authorversion]{acmart}
\AtBeginDocument{%
  }

\copyrightyear{2022}
\acmYear{2022}
\setcopyright{acmlicensed}\acmConference[ARES 2022]{The 17th International Conference on Availability, Reliability and Security}{August 23--26, 2022}{Vienna, Austria}
\acmBooktitle{The 17th International Conference on Availability, Reliability and Security (ARES 2022), August 23--26, 2022, Vienna, Austria}
\acmPrice{15.00}
\acmDOI{10.1145/3538969.3544430}
\acmISBN{978-1-4503-9670-7/22/08}

\begin{document}

\title{TaxIdMA: Towards a Taxonomy for Attacks related to Identities}

\author{Daniela Pöhn}
\email{daniela.poehn@unibw.de}
\orcid{0000-0002-6373-3637}
\author{Wolfgang Hommel}
\email{wolfgang.hommel@unibw.de}
\affiliation{%
  \institution{Universität der Bundeswehr München, RI CODE}
  \city{Munich}
  \country{Germany}
}

\renewcommand{\shortauthors}{Pöhn and Hommel}

\begin{abstract}
Identity management refers to the technology and policies for the identification, authentication, and authorization of users in computer networks. Identity management is therefore fundamental to today's IT ecosystem. At the same time, identity management systems, where digital identities are managed, pose an attractive target for attacks. With the heterogeneity of identity management systems, every type (i.\,e., models, protocols, implementations) has different requirements, typical problems, and hence attack vectors. In order to provide a systematic and categorized overview, the framework Taxonomy for Identity Management Attacks (TaxIdMA) for attacks related to identities is proposed. The purpose of this framework is to classify existing attacks associated with system identities, identity management systems, and end-user identities as well as the background using an extensible structure from a scientific perspective. The taxonomy is then evaluated with eight real-world attacks resp. vulnerabilities. This analysis shows the capability of the proposed taxonomy framework TaxIdMA in describing and categorizing these attacks.
\end{abstract}



\maketitle

\section{Introduction}

In order to provide access to several different services, organizations operate identity management systems (IdMS). These systems manage users with their digital identities, associated user information, i.\,e., attributes, authentication methods, and permissions. One often used IdMS is Microsoft Active Directory (AD), which does not only manage users but also devices. These systems typically come with other mature technologies including single sign-on and role-based access control. Attacks targeting identities often focus on IdMS as the central source of truth~\cite{mci/Fritsch2020}. However, security among other things relies on the correct setup and configuration. According to the Purple Knight Report 2022~\cite{purpleknight}, organizations scored an average of 68\% across five AD security categories, while large organizations achieved an even lower rating. This shows that organizations are challenged by securing AD, especially in larger environments. As industries move forward with outsourcing, the reliance on the correct implementation of processes and human factors becomes more problematic.

In addition, users can enable as well as be the target of attacks. For example, users tend to reuse or create simple passwords~\cite{10.1145/3183341}. Consequently, different brute-force attacks such as credential stuffing are possible, resulting in account takeover and, thereby, the loss of personal data. Last but not least, identities are used in every system, e.\,g., to run a web service. These identities are misused by attackers during the attack life cycle~\cite{strom2018mitre}. This demonstrates that identities are an important part of IT and, thereby, IT security. In this context, taxonomies provide an overview of the systems and different possibilities. Such a systematic approach helps to identify gaps, improve the current situation, and provide a guideline for establishing new security measures.

The contribution of the paper focuses on Taxonomy for Identity Management Attacks (TaxIdMA), which is a novel taxonomy framework for attacks related to identities. The following taxonomies are proposed: i) attack background, ii) attacks involving system identities, iii) attacks on IdMS, and iv) attacks on end-user identities. The taxonomy on the attack background describes the general setting of an account. The other taxonomies detail attacks and can be combined to categorize the different steps within an attack life cycle. Thereby, the taxonomies classify various attacks in greater detail than existing generic taxonomies. This can help to, e.\,g., improve the security of identities and systems, analyze incidents, and design new approaches. TaxIdMA is then evaluated based on eight real-world examples.

The remainder of the paper is as follows: Section~\ref{sec:sota} discusses existing taxonomies and categorizations. Section~\ref{sec:TaxIdMA} describes the proposed taxonomies of TaxIdMA. Section~\ref{sec:evaluation} evaluates the proposed taxonomies by applying real-world examples. Based on the evaluation, Section~\ref{sec:discussion} discusses the benefits and limits of the taxonomies. Last but not least, Section~\ref{sec:conclusion} concludes the paper and gives an outlook on future work.

\section{Related Work}
\label{sec:sota}

Different criteria and taxonomies group incidents involving identities. Common Weakness Enumeration (CWE) by MITRE~\cite{cwe} is a community-developed list of software and hardware weaknesses. This list serves as a common language for weakness identification, mitigation, and prevention efforts. The research concept of improper access control includes, e.\,g., different types of improper access controls ranging from Hypertext Transfer Protocol (HTTP) cookies to on-chip hardware issues. Other categories also contain weaknesses related to identity management. Common Attack Pattern Enumerations and Classifications (CAPEC)~\cite{capec} relates to CWE and Common Vulnerabilities and Exposures (CVE). Here, attack patterns are based on software design patterns. Subvert of access control includes, e.\,g., authentication abuse and bypass as well as physical theft, without including all issues related to it. Further categorizations have a section about identity management without going into details. Open Web Application Security Project (OWASP)~\cite{owasp} publishes top 10 lists and cheat sheets for typical problems.

Igure and Williams~\cite{4483667} analyze taxonomies published from 1974 until 2006 resulting in a taxonomy of attacks and vulnerabilities in computer systems. Even though the authors include numerous approaches, the proposed taxonomy is generic as it is not focused on a specific security area. Chapman et al.~\cite{10.5555/2048558.2048569} propose a 3-tier taxonomy to describe the effects of cyber attacks, ranging from no access to user access to root access requirements. The authors thereby characterize the different levels of permissions an attacker gained. Nevertheless, it is unclear whether service users like \texttt{www-data} are included in user access or nowhere. Derbyshire et al.~\cite{8406575} evaluate different taxonomies based on a criteria set and real-world attacks. CAPEC outperforms the other taxonomies, though the different versions and criteria might be challenging. Furthermore, the authors notice that several taxonomies do not include human elements. Cho et al.~\cite{8551383} explore cyber kill chain models resulting in a new model to evaluate cyber situation awareness. Other taxonomies have a specific focus. Habiba et al.~\cite{Habiba2014} propose a taxonomy for security issues related to cloud IdMS. Klaper and Hovy~\cite{10.1145/2612733.2612759} create a basic taxonomy with cybersecurity topics linked to relevant educational or research material. Burger et al.~\cite{10.1145/2663876.2663883} suggest a cyber threat intelligence information exchange taxonomy. Husseis et al.~\cite{8888436} provide a review of potential threats affecting biometric systems. Williams et al.~\cite{8884913} classify security features in Internet of Things (IoT) devices.

As a result, a holistic taxonomy for attacks related to identities is still missing.

\section{TaxIdMA}
\label{sec:TaxIdMA}

In this section, the framework TaxIdMA with its taxonomies is described. A taxonomy organizes its concepts hierarchically by following these ideal properties~\cite{10.1145/185403.185412,601330}: i) mutually exclusive categorization, i.\,e., no overlapping, ii) clear and unambiguous classification criteria, and iii) comprehensible, useful, and compliant with established terminology. TaxIdMA was generated through a regression manner -- abstraction of knowledge from existing taxonomies, other related approaches, and known attacks. During the design, attention was paid to the complete coverage of all necessary factors of the attacks. The goal of TaxIdMA is to categorize all steps of an attack related to identities.

TaxIdMA consists of the taxonomies attack background, system identities, identity management systems, and end-user identities. It uses the definitions summarized in Appendix~\ref{sec:definitions}. The taxonomy on attack background, described in Section~\ref{sec:attackbackground}, can be used for all attacks involving identities as it outlines the background of the attack. The background is constant during the attack cycle, which can involve several identities. Categories with unmodified values are grouped there; categories, where the values may vary either depending on the attack type or during the attack cycle are listed in the following taxonomies. The taxonomies on system identities, identity management systems, and end-user identities further detail the attack. During the attack cycle, the attacker typically reuses system identities. This circumstance is described in the related taxonomy, see Section~\ref{sec:systemidentities}. If multiple system identities are utilized, the taxonomy can be applied several times.
One major goal of attacks are IdMSs, as all accounts are managed there. Hence, it is described in another taxonomy. With the outsourcing of services including IdMS, several entities may be involved. A categorization of the steps towards an attack is shown in the according taxonomy in Section~\ref{sec:identitymanagementsystems}. Last but not least, another kind of attack targets end-user identities, either in rather selected attacks, like spear-phishing, or in massive attacks, e.\,g. broader-scale phishing. As they focus on end-users and not organizations, another taxonomy is established in Section~\ref{sec:enduseridentities}.

Considering that an attacker may exploit various identities, several up to all taxonomies can be applied in a stepwise way. For example, a spear-phishing attack targets an employee with the help of an identity leak for a service the employee uses, dropping malware, which then gives access to a system identity with the ultimate goal of the IdMS. In this example, all taxonomies are utilized to systematically describe the attack, either from the start to the end or vise versa.

\subsection{Attack Background}
\label{sec:attackbackground}

\begin{figure}[!htp]
\centering
\includegraphics[width=0.5\textwidth]{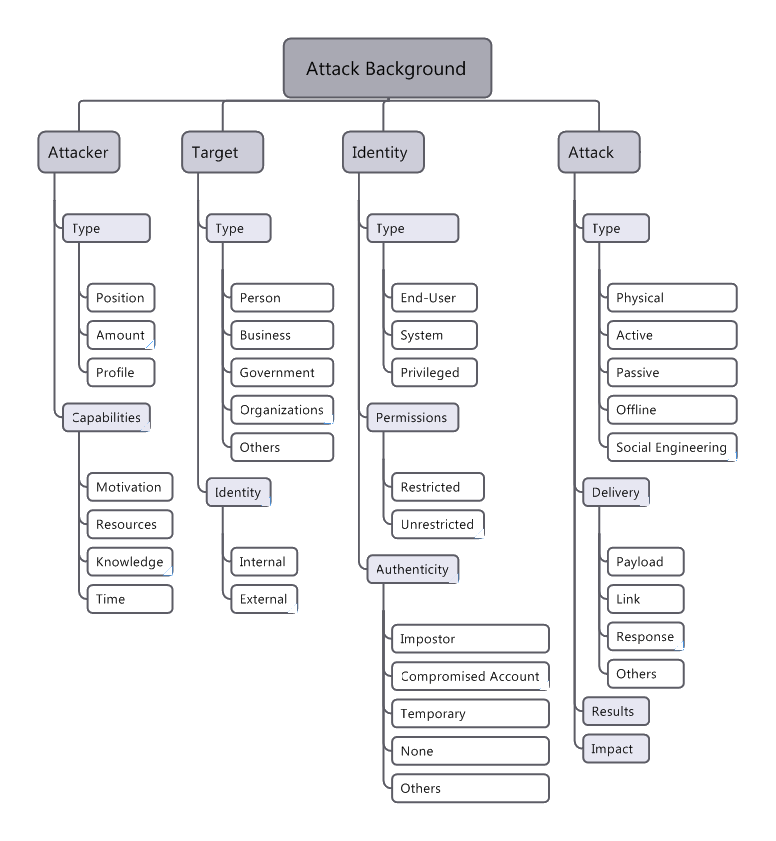}
\Description[<Taxonomy for attack background>]{<Taxonomy for attack background>}
\caption{Taxonomy for attack background}
\label{fig:background}
\end{figure}

The attack background taxonomy, shown in Figure~\ref{fig:background}, is categorized by the attacker, target, attack identity, and the attack itself. It describes the background of the attack, detailed by the different attack taxonomies in the following sections.

\textbf{Attacker:} Someone who explores methods for breaching weaknesses in a computer system or network. The attacker is detailed by type and capabilities~\cite{CHNG2022100167}.
\begin{itemize}
\item \emph{Type}: The type of attacker describes the position (i.\,e., internal or external), the number of involved persons (amount), and their profiles (single or groups such as cybercriminals or state-sponsored hackers).
\item \emph{Capabilities}: Expertise or ability of the attacker to reach the goal. The capabilities are characterized by motivation, resources, knowledge, and time. Determined by the capabilities, the severity of the attack varies.
\end{itemize}

\textbf{Target:} A goal designated for an attack. The target is described by identity, type, and sector~\cite{HANSMAN200531,simmons2014avoidit}.
\begin{itemize}
\item \emph{Type:} Attacks focus on different targets, ranging from single persons, businesses, governments, and organizations to other types. These can be grouped into sectors.
\item \emph{Identity:} The identity describes the position of the target related to the attacker and can be further detailed by their roles. The identity of the target is either internal or external. Internal can be characterized by the various types of identities, which have different permissions. As an example, executive, employee, administrator, and contractor are named. External contains partner, customer, trusted third party, and stranger.
\end{itemize}

\textbf{Identity:} Digital identity used during the attack with permissions~\cite{10.5555/2048558.2048569} and authenticity.
\begin{itemize}
\item \emph{Type:} Type of digital identity, i.\,e., end-user, system, and privileged~\cite{8551383}. In the beginning, the attacker typically has no identity.
\item \emph{Permissions:} Permissions (authorization) the overtaken identity and, therefore, the attacker has at the described moment. Identities come with permissions according to roles and functions, ranging from restricted to unrestricted~\cite{10.5555/2048558.2048569}.
\item \emph{Authenticity:} The authenticity of the attacker towards the system during the attack. The type of identity has one of the following authenticities: impostor (e.\,g., during phishing attacks), the authenticity of the compromised account (e.\,g., during a system attack), a temporary authenticity (e.\,g., if the attacker has the possibility to create accounts), none, or others. The authenticity was added to describe the human element~\cite{8406575}.
\end{itemize}

\textbf{Attack:} The use of an exploit by an adversary to take advantage of a weakness with the intent to achieve a negative impact. The attack is described by type, delivery, results, and impact~\cite{HANSMAN200531,simmons2014avoidit}.
\begin{itemize}
\item \emph{Type:} Characterization of the threat similar to CAPEC. Physical attacks contain, e.\,g., theft. Active attacks can again be divided and include, e.\,g., identity theft and interruptive attacks ((distributed) denial of service). Passive attacks describe eavesdropping and other passive methods. Offline attacks comprise of cracking attacks, password speculations, and crypto analyses. Last but not least, social engineering is named, which may include other types, e.\,g., passive (research), active (actual attack), and physical (e.\,g., tailgating). Even though it may contradict the outline, it was added to include human elements~\cite{8406575}.
\item \emph{Delivery:} Way of delivering the attack. This consists of payloads (e.\,g., a reverse shell), links (e.\,g., phishing links), responses (e.\,g., server or email responses), and others (e.\,g., physical)~\cite{10.1145/2835375}.
\item \emph{Results:} Direct consequences of an attack, ranging from nuisance, degradation, and disruption to disabling, theft, and disclosure.
\item \emph{Impact:} Loss or the consequences which are incurring due to the attack. For example, business disruption, intellectual property loss, customer information loss, reputation loss, and financial loss. Further details of results and impact were intentionally left out of the figure for clarity and can be seen in Figure~\ref{fig:background-ext} in Appendix~\ref{sec:appendix}.
\end{itemize}

\subsection{System Identities}
\label{sec:systemidentities}

\begin{figure}[!htp]
\centering
\includegraphics[width=0.5\textwidth]{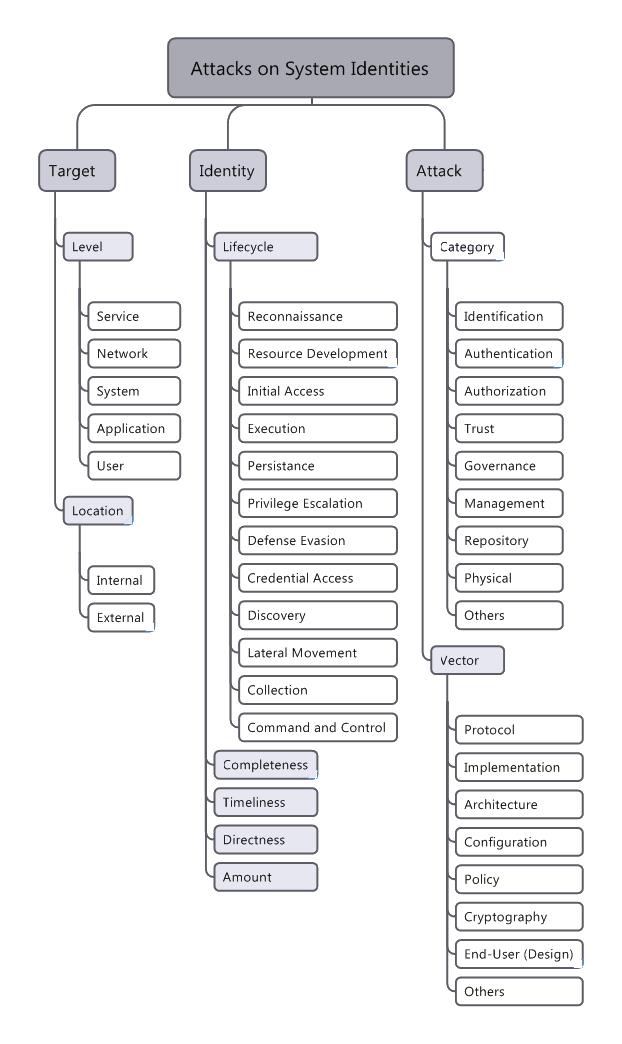}
\Description[<Taxonomy for attacks using system identities>]{<Taxonomy for attacks using system identities>}
\caption{Taxonomy for attacks using system identities}
\label{fig:system}
\end{figure}

The system identities taxonomy is relevant during almost all attacks (besides denial of service and eavesdropping), because attackers take over different digital identities to progress towards their goal. As shown in Figure~\ref{fig:system}, the taxonomy further details target, identity, and attack.

\textbf{Target:} In accordance with the taxonomy of the attack background, the target is further specified.
\begin{itemize}
\item \emph{Level:} Target level in the system stack. Identities appear on different levels: service, network, system with cryptography and hardware, applications with server (database, storage, web, email, etc.) and client as well as user~\cite{bsi}. As the level varies across the taxonomies, it is detailed within the specific taxonomies. Even though the differences in the presented taxonomies are rather small, extensions may require other levels.
\item \emph{Location:} Physical location of the target. With outsourcing and cloud infrastructures, the location of the target in relation to the attacker may vary, from local to external, e.\,g., a trusted third party. \cite{Habiba2014}
\end{itemize}

\textbf{Identity:} The identity category details lifecycle, completeness, timeliness, directness, and amount.
\begin{itemize}
\item \emph{Lifecycle:} Stage of the attack lifecycle, also known as cyber kill chain~\cite{8551383,strom2018mitre}.
\item \emph{Completeness:} Completeness of identity control takeover, i.\,e., fully or partly.
\item \emph{Timeliness:} Timeliness of identity control takeover, i.\,e., definitely temporary or until recovery.
\item \emph{Directness:} Direction of targeting, i.\,e., directly or indirectly.
\item \emph{Amount:} Amount of targeted identities. With system identities, single or selected identities are typically under the control of the attacker. This may change within the following taxonomies, where multiple identities can be gained. The latter values were left out of the figure for clarity and can be found in Figure~\ref{fig:systems-ext} in Appendix~\ref{sec:appendix}.
\end{itemize}

\textbf{Attack:} The attack is described by category and vector.
\begin{itemize}
\item \emph{Category:} Targeted weakness of identity management, i.\,e., identification, authentication, authorization, trust, governance, user management, user repository, physical or others~\cite{capec}.
\item \emph{Vector:} Specific path, method, or scenario exploited. This can be protocol, implementation, architecture, configuration, policy, crypto\-graphy, end-user design, and others.
\end{itemize}

\subsection{Identity Management Systems}
\label{sec:identitymanagementsystems}

\begin{figure}[!htp]
\centering
\includegraphics[width=0.5\textwidth]{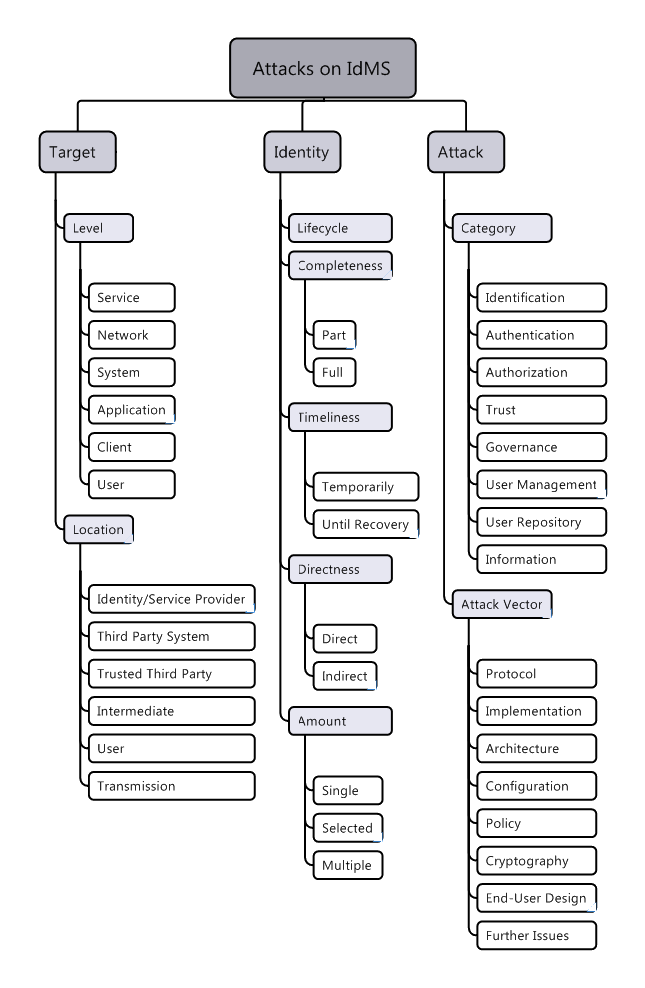}
\Description[<Taxonomy for attacks on identity management systems>]{<Taxonomy for attacks on identity management systems>}
\caption{Taxonomy for attacks on identity management systems}
\label{fig:idms}
\end{figure}

IdMSs are an important goal during attacks. As shown in Figure~\ref{fig:idms}, the location within targets contains more entities as IdMS may be used in cooperation with third parties. In addition, the attack vector is further divided and several examples are explained.

\textbf{Target:} The target is detailed by level and location.
\begin{itemize}
\item \emph{Level:} The level is similar to the taxonomy for system identities~\cite{bsi}, shown above.
\item \emph{Location:} As IdMS can be used across organizations (e.\,g., cross-organization IdM, IdM-as-a-Service), identity provider resp. service provider, third party system, trusted third party, intermediate, end-users, and transmission are possible~\cite{Habiba2014}.
\end{itemize}

\textbf{Identity:} Similarly to the taxonomy for system identities, the identity is described by the lifecycle, completeness, timeliness, directness, and amount.
\begin{itemize}
\item \emph{Lifecycle:} The attack can involve the IdMS in different stages within the lifecycle, even though the IdMS is typically the first main target~\cite{8551383,strom2018mitre}. The lifecycle is intentionally left out and shown in Figure~\ref{fig:idms-ext} in Appendix~\ref{sec:appendix}.
\item \emph{Completeness:} An identity and IdMS can partly or fully be taken over, see, e.\,g., silver tickets for AD.
\item \emph{Timeliness:} The timeliness can differ, although until recovery is standard for IdMS.
\item \emph{Directness:} An attack can either be direct or indirect.
\item \emph{Amount:} Typically, the number of targeted accounts is multiple, though it could be selected or single. The amount may increase during the attack.
\end{itemize}

\textbf{Attack:} An attack is detailed by category and vector. The vector is use case-specific and explained in the following.
\begin{itemize}
\item \emph{Category:} Similarly to system identities, the IdMS taxonomy uses a category of attacks~\cite{capec}. This can include identification, authentication, authorization, trust, governance, user management, user repository, and information.
\item \emph{Vector:} The attack groups are further divided into protocol, implementation, architecture, policy, crypto\-graphy, end-user design, and further issues. Following are two examples for implementation and configuration.
The implementation of AD had vulnerabilities including MS17-010 Eternal Blue and MS16-032 in earlier versions. The AD implementation of Kerberos could be used for Pass-the-Hash and Kerberoasting.

The configuration of AD has several pitfalls, grouped into account (e.\,g., password in comments), group (e.\,g., built-in groups and unlimited groups), and delegation. Depending on, for example, the configuration of the Lightweight Directory Access Protocol (LDAP) implementation or finger, enumeration is possible.
\end{itemize}

\subsection{End-User Identities}
\label{sec:enduseridentities}

\begin{figure}[!htp]
\centering
\includegraphics[width=0.5\textwidth]{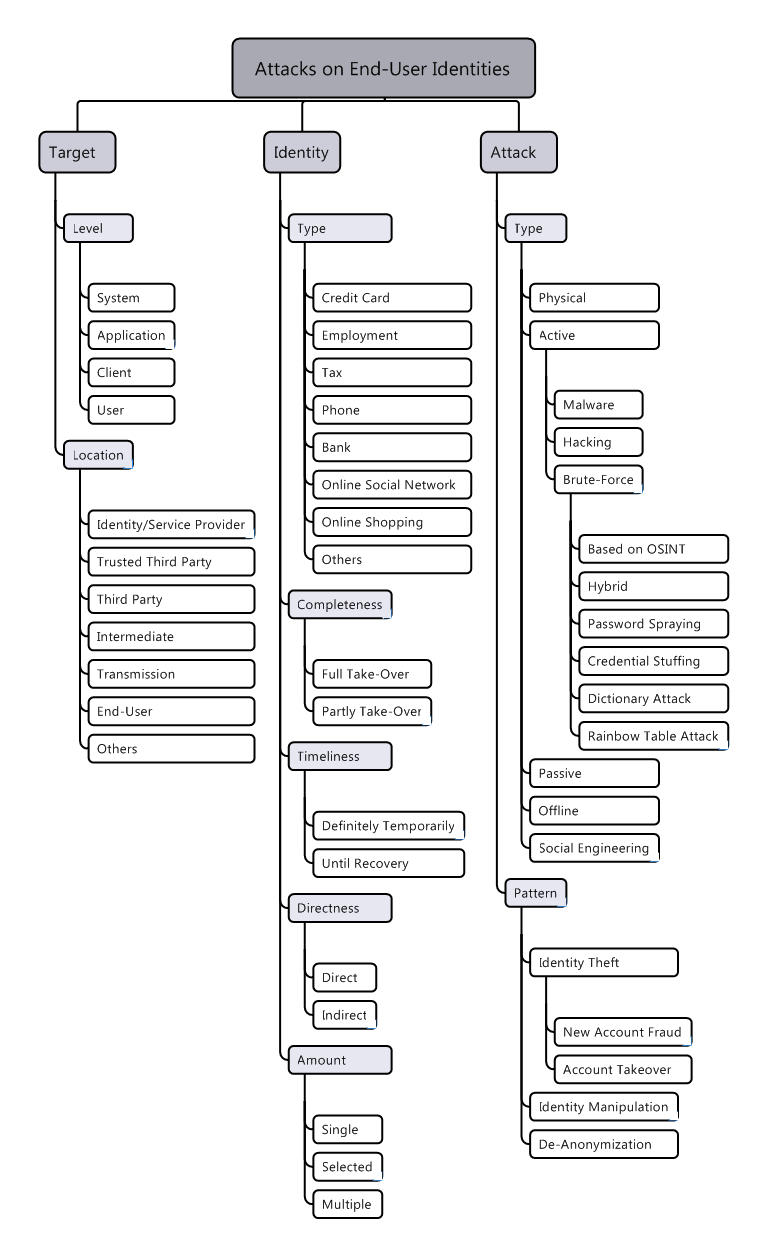}
\Description[<Taxonomy for attacks on end-user identities>]{<Taxonomy for attacks on end-user identities>}
\caption{Taxonomy for attacks on end-user identities}
\label{fig:identities}
\end{figure}

The end-user identities taxonomy is shown in Figure~\ref{fig:identities}. It focuses on end-user identities, which are attacked, either directly or indirectly. While an individual digital identity has little financial value, the amount makes these types of attacks powerful. Hence, the taxonomy includes additional identity types, while the number of targeted identities is mostly multiple. The type of attack is concretized, while an additional pattern is included. The attack pattern further details and describes the purpose of the attack.

\textbf{Target:} The target end-user limits the possibilities of type and identity. Instead, level and location are added.
\begin{itemize}
\item \emph{Level:} End-user identities appear on fewer levels: system, application, client, and user.
\item \emph{Location:} As identities are stored in databases and IdMS, the same locations are possible: identity resp.  service provider, trusted third party, third party, intermediate,  transmission, end-user, and others.
\end{itemize}

\textbf{Identity:} With the focus on end-users, the type shifts.
\begin{itemize}
\item \emph{Type:} The identity type related to end-users is different from the types described above. Typical types contain information resp. accounts about credit card (including child credit card history), employment, tax, phone, bank, online social networks, online shopping, and other accounts.
\item \emph{Completeness:} Completeness is divided into full and partial take-over. Partial take-over is, e.\,g., possible during session hijacking.
\item \emph{Timeliness:} Timeliness is defined as either definitely temporarily or until recovery. The session is hijacked as long as the session exists. Therefore, the timeliness of the example is temporarily.
\item \emph{Directness:} The attack can either be direct or indirect.
\item \emph{Amount:} While the data from one end-user identity is cheap, the number of compromised accounts makes attacks valuable. Nevertheless, the attacker may target selected and single identities as well.
\end{itemize}

\textbf{Attack:} Also here, the types change and patterns further define the attack.
\begin{itemize}
\item \emph{Type:} The attack type contains the same categories as the taxonomies beforehand. Passive may be the underground economy, open-source intelligence (OSINT), eavesdropping, and cracking. Physical especially focuses on things and devices. Another category is social engineering such as phishing. Within active attacks, malware (e.\,g., keylogger), hacking, and brute force are possible. Brute force related to identities includes password spraying, credential stuffing, dictionary attacks, rainbow table attacks, attacks based on OSINT information, and hybrid mode. \cite{capec}
\item \emph{Pattern:} Description of the methodology used by the adversaries to exploit weaknesses. The attack pattern consists of identity theft, identity manipulation, and de-anonymization. Identity theft is further divided into new account fraud (e.\,g., existing profile cloning attack) and account takeover. Both can be combined. This category relates to CAPEC, CWE, and OWASP.
\end{itemize}

\section{Evaluation of TaxIdMA}
\label{sec:evaluation}

In this section, the taxonomy framework TaxIdMA is evaluated. First, the applied methodology is described, followed by the evaluation.

\subsection{Methodology}

\renewcommand*{\arraystretch}{1.1}
\begin{table*}[!htp]
\caption{Overview of the selected examples used within the evaluation\label{tab:evaluation}}
\begin{tabular}{l|llll}
                                    & \textbf{Background} & \textbf{System} & \textbf{IdMS} & \textbf{End-User} \\ \hline
\textbf{Vulnerabilities/Attacks}    & 1/7                 & 1/1             & 0/3           & 0/3               \\
\textbf{APT}                        & 1                   & 0               & 1             & 0                 \\
\textbf{Supply Chain Attack}                        & 1                   & 0               & 1             & 0                 \\
\textbf{Web Application/Smartphone} & 3/1                 & 1/0             & 0/0           & 2/1               \\
\textbf{Computer/Server}            & 1/3                 & 1/0             & 0/3           & 0/0               \\
\textbf{Year}                       & 2014-2022           & 2022            & 2020, 2022    & 2014, 2021, 2022 \\ \hline
\textbf{Amount}                     & 8                   & 2               & 3             & 3                 \\
\end{tabular}
\end{table*}
\renewcommand*{\arraystretch}{1}

In order to analyze TaxIdMA, we apply eight real-world examples. These include one vulnerability, a smartphone app, and an advanced persistent threat, which is at the same time a supply chain attack. The examples were selected so that they are as diverse and up-to-date as possible. There is one exception: Celebgate already happened in 2014. Nevertheless, it can be categorized. All selection criteria are shown in Table~\ref{tab:evaluation}.

\subsection{Evaluation}

First, the vulnerability or incident is described, followed by the categorization of the background. Last but not least, the mapping to the detailed taxonomy is explained. The first example, Zoom ZSB-22004, additionally highlights the categorization in the respective taxonomies.

\subsubsection{Zoom ZSB-22004}

Zoom revealed the local privilege escalation vulnerability ZSB-22004 (Common Vulnerabilities and Exposures
(CVE)-2022-22782) in April 2022 with a high Common Vulnerability Scoring System (CVSS) score of 7.9. During the installer repair operation, malicious actors could utilize the vulnerability for further actions. \cite{zoom}

\begin{figure}[!htp]
\centering
\includegraphics[width=0.5\textwidth]{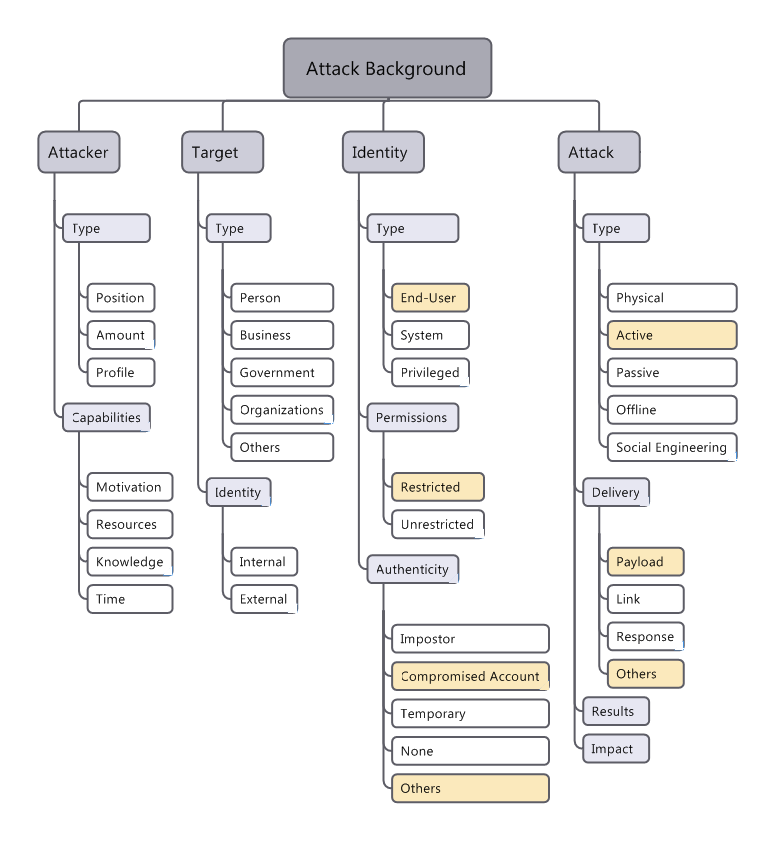}
\Description[<Taxonomy for attack background used for Zoom>]{<Categorization of Zoom ZSB-22004 based on the taxonomy for attack background>}
\caption{Categorization of Zoom ZSB-22004 based on the taxonomy for attack background}
\label{fig:background-zoom}
\end{figure}

First, the attack background is described based on Figure~\ref{fig:background-zoom}. As it is a vulnerability, the categories ``Attacker'' and ``Target'' cannot be further specified. For the category ``Identity'', the following values are selected: The \emph{type} end-user identity with limited \emph{permissions} is used. The attacker's \emph{authenticity} can either be the end-user itself or utilize the compromised account. Within the category ``Attack'', we see the following: The attack has the \emph{type} active and might be helped with social engineering. The \emph{delivery} depends on the actual attack and is either payload or others. The \emph{result} is at least escalated privileges, while the \emph{impact} varies.

\begin{figure}[!htp]
\centering
\includegraphics[width=0.5\textwidth]{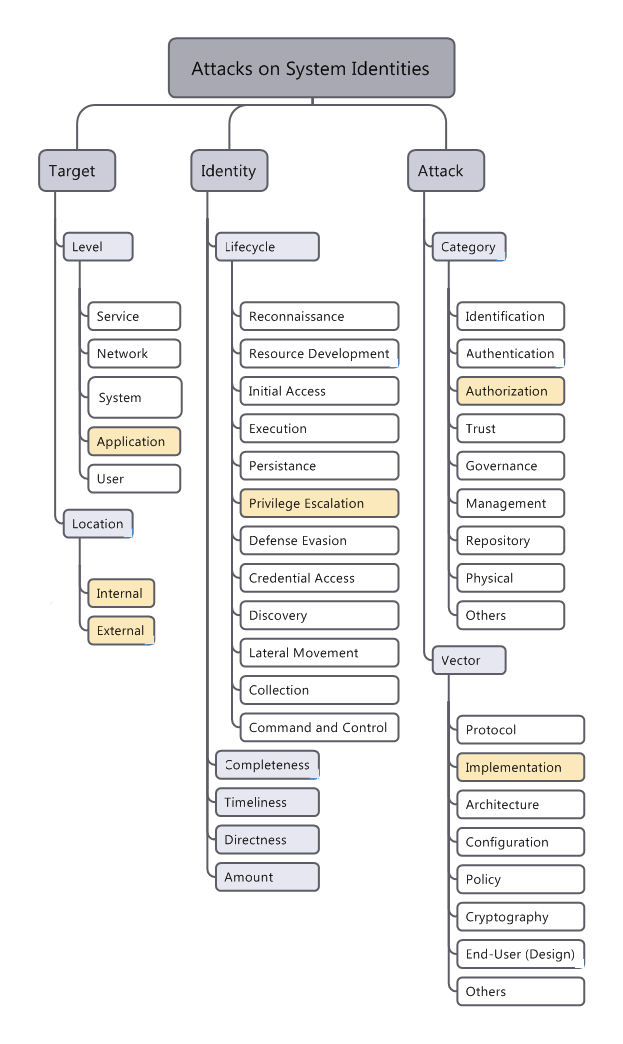}
\Description[<Taxonomy for attacks using system identities used for Zoom>]{<Categorization of Zoom ZSB-22004 based on the taxonomy for attacks using system identities>}
\caption{Categorization of Zoom ZSB-22004 based on the taxonomy for attacks using system identities}
\label{fig:system-zoom}
\end{figure}

As the vulnerability can be used to escalate privileges at an end-user system, we select the taxonomy related to system identities shown in Figure~\ref{fig:system-zoom}. For the category ``Target'', we detail our description with the following: The vulnerability affects the application \emph{level} and could have consequences on the \emph{location}, which is either internal or external. The vulnerability may have the following ``Identity'': It is used during privilege escalation within the \emph{lifecycle}. If the attacker manages to establish persistent access, then the \emph{completeness} and \emph{timeliness} are given, otherwise not. The \emph{amount} is single to selected and the identities have the \emph{directness} directly. Last but not least, the category ``Attack'' is specified. The attack targets the \emph{category} authorization and uses the \emph{vector} implementation of Zoom.

\subsubsection{VirusTotal}

VirusTotal is a web application with more than 70 antivirus scanners that scan files and Uniform Resource Locators (URLs) sent by users. Researchers of the CySource Team~\cite{virustotal} used the platform to upload a payload, which was added to the file's metadata, to the different hosts to perform the scan. When the \texttt{exiftool} on the hosts was utilized, the uploaded payload was actually executed, giving the researchers reverse shells from more than 50 internal hosts with high privileges. Due to the unauthorized access, it was possible to obtain sensitive and critical information, such as tokens and certificates.

The attacker \emph{type} researcher is rather special as they want to play out of interest, typically work in teams, and are mostly outsiders. The \emph{capabilities} were more than script kiddie as they had a specific CVE (CVE-2021-22204) in mind and the knowledge to use it. The target \emph{type} was a selected service hosted by Google, which has the \emph{identity} external. The identity utilized had the \emph{type} end-user, while they gained privileged access. Thereby, they had almost unrestricted \emph{permissions} with the \emph{authenticity} of the compromised account. The attack \emph{type} was active (i.\,e., hacking), using a payload for \emph{delivery}, which was inserted into the file's metadata. The \emph{result} was the compromise of several servers and the leakage of several sensitive and critical information. The \emph{impact} may be press for the researchers as well as the service.

The target \emph{level} is system via applications and, therefore, has an external \emph{location}. Regarding the \emph{lifecycle}, the initial access was mostly also the privilege escalation, providing privileged access. While the researchers \emph{completely} took control and the \emph{timeliness} is until recovery, the \emph{directness} is indirect as they used the web application to get access. The \emph{amount} of gained identities are multiple. The main \emph{vector} is the implementation of \texttt{exiftool}, although there should be at least a policy in place to update vulnerable software. Probably the configuration (too many permissions) could be improved as well. Hence, the \emph{category} is others.

\subsubsection{Solarwinds from FireEye's Point of View}

In December 2020, FireEye reported an incident -- the first hint of the Solarwinds Hack~\cite{cisa}. A new registered phone of a FireEye user account raised a standard alarm. As it was the second enrolled phone for the user, the person was contacted and asked if it is their new phone. Even though it was a severity zero alert, an attacker tried to register a new device to bypass multi-factor authentication. Throughout investigations, the earliest trace of compromise was on the system with the Solarwinds product Orion. During the first phase of the attack, a backdoor was installed via Orion. In the second phase, this backdoor was actively used to collect domain credentials. The following phase involved the token signature certificates to access Office365. The last phase targeted the red team tools of FireEye. The attack does not solely concentrate on FireEye, but we take the company's point of view.

The attacker (\emph{type}) likely worked in a group with a high degree of \emph{capabilities}. The attack targeted the \emph{type} businesses, but also governments, and other organizations. The attacker had the \emph{identity} externals. During the attack, the attackers gained several identities with the \emph{type} ranging from end-user to privileged. As far as understood, they used compromised accounts with the respective \emph{permissions} and \emph{authenticity}. The attack was of an active \emph{type}, \emph{delivered} by the malicious code. The \emph{result} includes compromised software and networks, while the \emph{impact} is still not fully known.

One goal during the attack was the IdMS. Therefore, the target was \emph{located} at the identity provider for the \emph{level} systems and networks. The takeover was in a later stage of the \emph{lifecycle} with the \emph{timeliness} until recovery, \emph{directness} direct, and \emph{amount} selected accounts. The registration of the new phone focused on authentication. The attack \emph{vector} is others, as the attacker used the supply-chain attack with the software update of Orion.

\subsubsection{NVIDIA}

NVIDIA Corp. had an incident that completely compromised the company's internal systems over a week in February 2022. According to reports, the attackers got hold of an employee's account. The Twitter user vx-underground tweeted that LAPSUS\$ has claimed responsibility, leaking password hashes for NVIDIA employees and NVIDIA's official code signing certificates, indicating other data~\cite{nvidia}. According to LAPSUS\$, NVIDIA tried to hack back and ransomed their machines.

The attacker group LAPSUS\$ with certain \emph{capabilities} typically targets the \emph{type} large enterprises, i.\,e., the \emph{identity} external. The attackers preannounce their attacks and probably use different \emph{types} of identities, such as compromised accounts with varying \emph{permissions} and \emph{authenticity}, during the attack. The attack might have included the \emph{types} hacking and social engineering. As no more information was released, it is difficult to estimate \emph{delivery}. The \emph{results} and \emph{impact} might not fully be visible yet; at least the data was leaked.

The \emph{location} at the target is the identity provider, as the IdMS was involved. Hence, the \emph{amount} of multiple accounts of employees were leaked, resulting in fully \emph{complete} and \emph{timeliness} until recovery. The attack targeted the \emph{category} user management among others. Further categorization is not possible due to the limited information.

\subsubsection{ARcare}

The US healthcare provider ARcare~\cite{arcare} experienced a data security incident impacting computer systems and temporarily disrupting its services while affecting 345,353 individuals. The investigation concluded in March 2022 that an attacker had access to the ARcare's network over a five-week period. The attack involved a malware infection. No misuse was reported, but the potentially exposed data include names, social security numbers, drivers' licenses or state identification numbers, financial account information, and health information. A health record is worth more than complete credit card information or social security numbers.

The attacker has some \emph{capabilities} to target a specific business, ARcare, (\emph{type}) with probably an external \emph{identity}. As the data was accessed, the gained identity (\emph{authenticity} compromised or temporary account) had enough \emph{permissions}. Therefore, it was likely of the \emph{type} system or privileged. The attack was of an active \emph{type}, maybe \emph{delivery} with payload, and had the \emph{result} of malware infection and data loss. The \emph{impact} was business disruption and may include identity theft, which can have consequences on the business.

Both, system identities and IdMS, can be regarded. We concentrate on the customer data. The target is further detailed with \emph{level} network and system as well as the \emph{location} identity provider. The data was probably one of the goals of the attack and belongs to the \emph{lifecycle} steps credential access resp. data exfiltration. The \emph{amount} multiple access to data was fully (\emph{completeness} and \emph{timeliness}) gained. The attack targeted the \emph{category} user management, though it stays unclear which \emph{vector} enabled the attack.

\subsubsection{Celebgate}

An attacker managed to phish several targets to gain access to at least 50 Apple's iCloud and 72 Google accounts in 2014, stealing private photos and videos of celebrities. Partly fallback authentication with security questions and guessable answers was exploited. In some cases, the attacker utilized a software program to download the entire content of the target's iCloud backup. \cite{10.1007/978-3-319-18609-2_4}

The single attacker (\emph{type}) had the \emph{capability} and the time to write phishing mails and guess security questions with the motivation to steal photos and videos of mainly female celebrities. The attacker used the \emph{authenticity} of the service provider, gaining access to the \emph{type} end-user identities with their \emph{permissions}. The attack \emph{type} was social engineering (and partly brute-force based on OSINT) with the \emph{delivery} mode link, getting to the \emph{result} of stolen data and the \emph{impact} of the publication of several nude photos (though the connection to the publication of the photos was not found).

The target \emph{level} is the user with the \emph{location} identity provider as the data was stored there. The gained identities were accounts for online services (\emph{type}), which were fully gained until recovery (\emph{completeness} and \emph{timeliness}). The attacker \emph{directly} targeted these selected accounts (\emph{amount}). The attack mainly used the \emph{type} social engineering, though brute-force based on OSINT was successful for some accounts. The \emph{pattern} is identity theft with account takeover.

\subsubsection{Spotify Credential Stuffing}

In February 2021, Bob Diachenko uncovered that Spotify suffered from a second credential stuffing attack involving an estimated hundred thousand accounts~\cite{spotify}. In November 2020, researchers found a misconfigured and open Elasticsearch cloud database containing more than 3,890 million records. Both databases were owned by malicious third parties. The usernames and passwords might be from previously reported breaches or collections of data, which were reused against Spotify accounts.

The attackers have the \emph{capability} to operate Elasticsearch, but not enough to configure it securely. The motivation might be money if they are able to gain access to high-level accounts or sell them in large quantities. The target consists of the \emph{type} persons, i.\,e., \emph{type} end-users with limited \emph{permissions}. The \emph{authenticities} towards Spotify are the compromised accounts. The attack \emph{type} is active, possible by the \emph{delivery} responses. The attackers got hold of credentials of valid users (\emph{result}), while the \emph{impact} may range from nuisance to financial, depending on the further actions.

The target consists of the \emph{level} end-users where the credentials are stored at the \emph{location} identity provider. The credentials might be used for third parties. The stolen identities are from the \emph{type} online accounts. Each account is \emph{completely} and \emph{directly} taken over until recovery (\emph{timeliness}, e.\,g., password change and maybe enabling multi-factor authentication). The \emph{type} brute-force (i.\,e., credential stuffing) attack with the \emph{pattern} of  account takeover had the goal to gain the \emph{amount} multiple accounts.

\subsubsection{FlyTrap}

The Android malware FlyTrap spread to more than 10,000 targets through Google Play Store and third-party app marketplaces. According to Zimperium~\cite{flytrap}, nine bad apps acted as FlyTrap. The targeted users were told to log in with their Facebook accounts to cast their vote or collect a coupon code.  The Trojan hijacked the sessions and took over Facebook accounts. The app then got details like Facebook ID, location, email address, IP address, and cookies resp. tokens associated with the Facebook account, before abusing the Facebook account to spread the malware through personal messaging with links to the Trojan.

The attacker had enough \emph{capabilities} to build these apps and run the associated command and control (C2) server. The exact \emph{type} of attacker is not known. The target \emph{type} is end-users, resulting in the identity \emph{type} end-user identities with restricted access (\emph{permissions}). The \emph{authenticity} of the attacker is an imposter regarding the app and the compromised account when looking at the social engineering techniques used. The attack \emph{type} is active combined with social engineering. The \emph{delivery} was through the app (i.\,e., others) and link. FlyTrap \emph{resulted} in several overtaken accounts and sessions stored in the command and control server. The \emph{impact} might increase if a vulnerability of the C2 server is exploited.

The attack targeted end-users via applications (\emph{level}). While the identity data is stored at the identity provider (\emph{location}), the data was gained via the aimed users resp. due to third parties. The goal (\emph{type}) was the partly and temporarily take over (\emph{completeness} and \emph{timeliness}) of Facebook social network accounts. The malware \emph{directly} targeted the accounts while the \emph{amount} is multiple. The attack \emph{type} is active malware with the \emph{pattern} account takeover.

\section{Discussion}
\label{sec:discussion}

During the design of the taxonomies, several iterations were made. Whereas the background is based on the common description of an attack, the other taxonomies are to a greater extent based on attacks and related work. One evolutionary iteration enabled the description of a stepwise attack, where the background is static while the identities of the attacker and, thereby, taxonomies change. Next, the categories were aligned across all taxonomies. Items, where the values may change, were added to the specific taxonomies, whereas items with static values were appended to the background.

A taxonomy should fulfill the requirements described in Section~\ref{sec:TaxIdMA}. The categories are mutually exclusive and, therefore, not overlapping. The classification criteria are (mostly) clear and unambiguous. While this is true for involved experts, a to-be-established tutorial may open the taxonomy to a wider audience. In addition, social engineering attacks may combine further attack types. Therefore, another iteration may be needed in the future. The background can be seen as the basis, whereas the other three parts of TaxIdMA detail the taxonomies for the specific use case. If an IdMS is targeted, the system identity taxonomy can be used for the steps beforehand. Depending on the attack, a taxonomy could be used several times to specify the explicit steps. Also here, a tutorial would provide added value for the wider audience. The evaluation showed that categorizing eight real-world and typical attacks is possible without difficulties. In future work, further attacks and typical problems can be analyzed. The taxonomy is comprehensive and uses established terminology in the fields of identity management and cyber security.

Even though the selected examples were easy to categorize and provide more details than existing approaches, the taxonomy is nevertheless rather generic concerning the variants of attacks focusing on AD. IoT devices can be explained by the system and end-user identities, while the management is again IdMS. A taxonomy specifically for IoT may help to show the variations and secure the systems. With the evolution of IT, further parts of the taxonomy may be needed in the future. The proposed TaxIdMA nevertheless provides a solid basis.

\section{Conclusion and Outlook}
\label{sec:conclusion}

Identity management is essential for the operation of all IT services. At the same time, IdMSs pose important targets for attacks~\cite{mci/Fritsch2020}. Therefore, their design, implementation, and configuration are crucial for the secure operation of the infrastructure. In order to provide a common ground, TaxIdMA was established. The taxonomy framework for identity management consists of several taxonomies: general attack background, system identities, IdMS, and end-user identities. The proposed taxonomy framework is a start towards a well-defined taxonomy for attacks focusing on identities. The evaluation of TaxIdMA with eight real-world examples showed that the taxonomy helps to categorize attacks and hence identify critical elements. However, with new developments, TaxIdMA may need improvements and refinements. In future work, we plan to evaluate more attacks and common problems to further detail the taxonomy. A tutorial will make TaxIdMA useful for a wider audience. Last but not least, defense mechanisms will be mapped to the attacks and grouped in a further taxonomy framework.

\begin{acks}
This work is partly funded by the Bavarian Ministry for Digital Affairs (Project DISPUT/STMD-B3-4140-1-4). The authors alone are responsible for the content of the paper.
\end{acks}

\bibliographystyle{ACM-Reference-Format}
\bibliography{TaxIdM}

\newpage
\appendix

\section{Taxonomies}

This appendix defines the terms used in the taxonomies and visualizes the extended taxonomies.

\subsection{Definitions}
\label{sec:definitions}

In the following, definitions for the terms used within the taxonomies are given.

\begin{itemize}
\item \textbf{Amount:} Quantity of targeted identities.
\item \textbf{Attack:} The use of an exploit by an adversary to take advantage of a weakness with the intent to achieve a negative impact.
\item \textbf{Attack Category:} Targeted weakness of identity management.
\item \textbf{Attack Pattern:} Description of the methodology used by the adversaries to exploit weaknesses.
\item \textbf{Attack Vector:} Specific path, method, or scenario exploited.
\item \textbf{Attacker:} Someone who explores methods for breaching weaknesses in a computer system or network.
\item \textbf{Authenticity:} Attribution of the attacker towards the system during the attack.
\item \textbf{Capabilities:} Expertise or ability of the attacker to reach the goal.
\item \textbf{Completeness:} State or condition of being complete concerning the identity control takeover.
\item \textbf{Delivery:} Way of conveying the attack.
\item \textbf{Directness:}  State of direction of targeting.
\item \textbf{Identity:} Digital identity used during the attack.
\item \textbf{Impact:} Loss or the consequences which are incurring (effects) due to the attack.
\item \textbf{Lifecycle:} Stage of attack lifecycle, also known as cyber kill chain~\cite{8551383,strom2018mitre}.
\item \textbf{Permissions:} Authorization of the overtaken digital identity.
\item \textbf{Results:} Direct consequences (final product) of an attack.
\item \textbf{Target:} A goal designated for an attack.
\item \textbf{Target Level:} Target position in the system stack.
\item \textbf{Target Location:} Particular place in physical space of the target.
\item \textbf{Target Identity:} Position of the target related to the attacker.
\item \textbf{Timeliness:} State and duration of being timely concerning the identity control takeover.
\item \textbf{Type:} A grouping based on shared characteristics.
\end{itemize}

\subsection{Extended Taxonomies}
\label{sec:appendix}

\begin{figure*}[!hpb]
\centering
\includegraphics[width=0.5\textwidth]{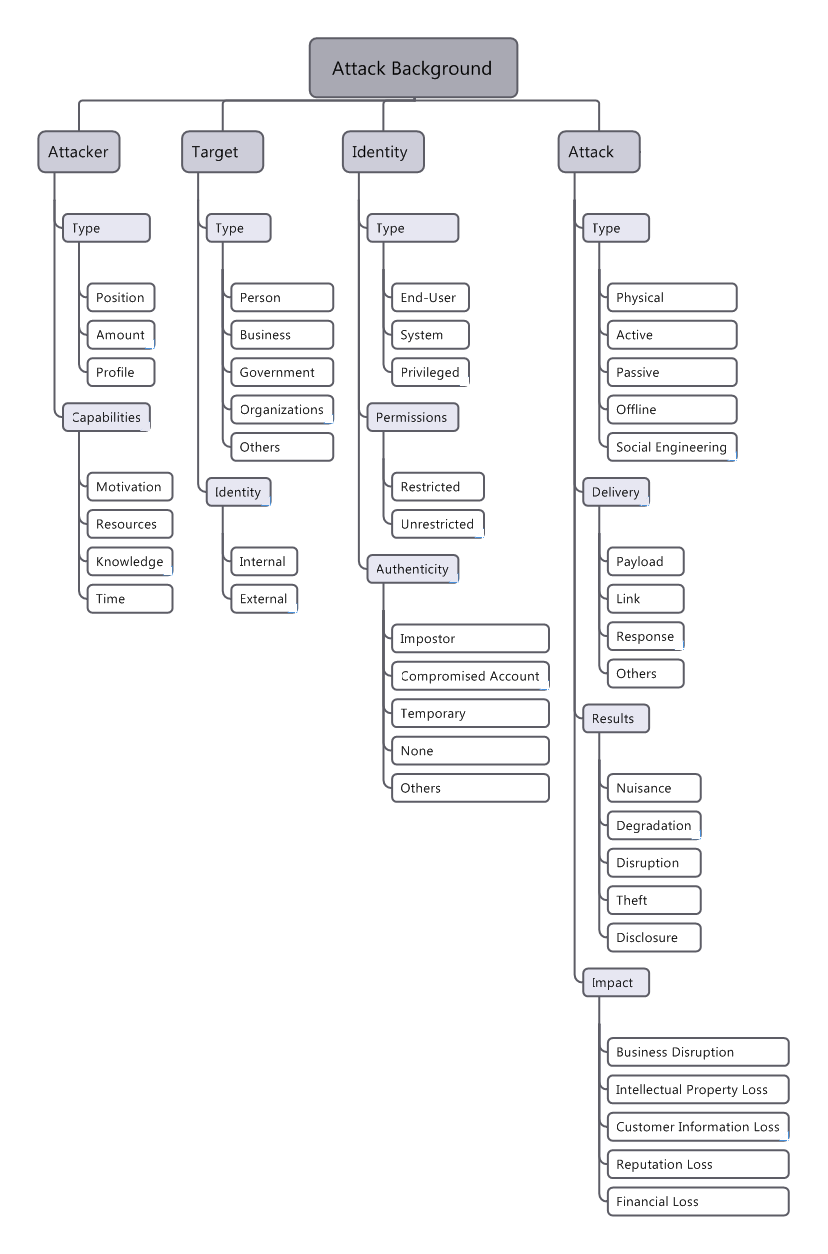}
\Description[<Extended taxonomy for attack background>]{<Extended taxonomy for attack background>}
\caption{Extended taxonomy for attack background}
\label{fig:background-ext}
\end{figure*}

\begin{figure*}[!htp]
\centering
\includegraphics[width=0.45\textwidth]{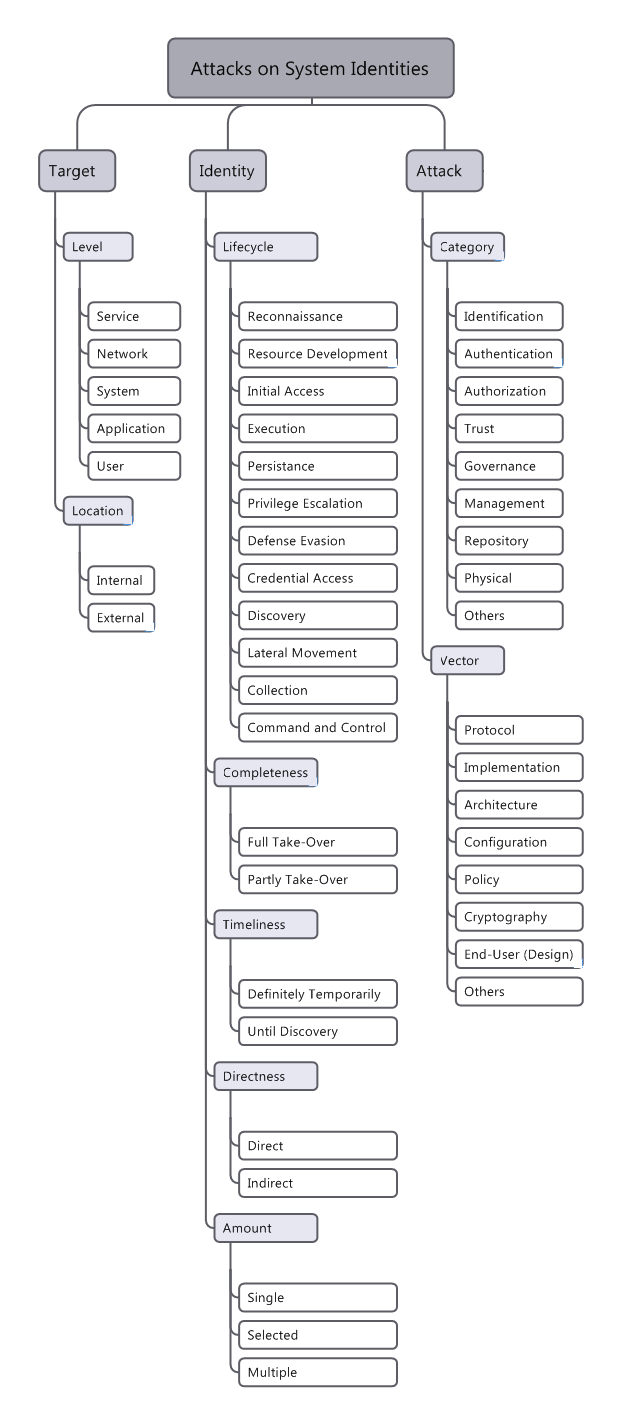}
\Description[<Extended taxonomy for attacks using system identities>]{<Extended taxonomy for attacks using system identities>}
\caption{Extended taxonomy for attacks using system identities}
\label{fig:systems-ext}
\end{figure*}

\begin{figure*}[!htp]
\centering
\includegraphics[width=0.5\textwidth]{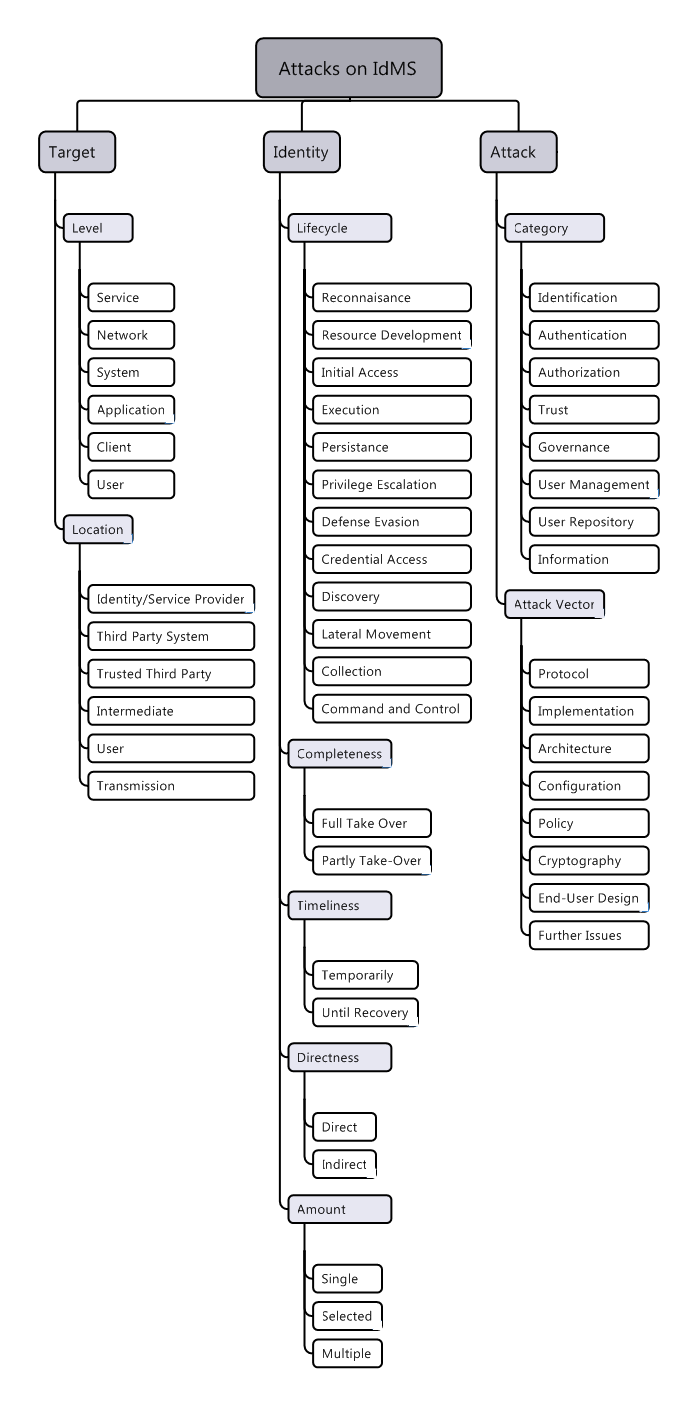}
\Description[<Extended taxonomy for attacks on identity management systems>]{<Extended taxonomy for attacks on identity management systems>}
\caption{Extended taxonomy for attacks on identity management systems}
\label{fig:idms-ext}
\end{figure*}

In the following, the extended taxonomies for attack background (see Figure~\ref{fig:background-ext}), system identities (see Figure~\ref{fig:systems-ext}), and IdMS (see Figure~\ref{fig:idms-ext}) are shown.

\end{document}